\documentclass[aps,prd,preprint,showpacs,nofootinbib]{revtex4}
\usepackage{graphicx}
\usepackage{amsmath}
\usepackage{bbm}
\usepackage{bm}
\usepackage{epsfig}
\usepackage{amssymb}
\def\sl{\!\!\!/}
\begin{document}
\title{
Quark fragmentation into spin-triplet $\bm{S}$-wave quarkonium
}
% \altaffiliation{}
%%%%%%%%%%%%%%%%%%%%%%%%%%%%%%%%%%%%%%%%%%%%%%%%%%%%%%%%%%%%%%%%%%%%%%%%%%%%%%
\author{Geoffrey~T.~Bodwin}
\affiliation{High Energy Physics Division, Argonne National Laboratory,\\
9700 South Cass Avenue, Argonne, Illinois 60439, USA}
\author{Hee~Sok~Chung}
\affiliation{High Energy Physics Division, Argonne National Laboratory,\\
9700 South Cass Avenue, Argonne, Illinois 60439, USA}
\author{U-Rae~Kim}
\affiliation{Department of Physics, Korea University, Seoul 136-713, Korea}
\author{Jungil~Lee}
\affiliation{Department of Physics, Korea University, Seoul 136-713, Korea}

\date{\today}
%%%%%%%%%%%%%%%%%%%%%%%%%%%%%%%%%%%%%%%%%%%%%%%%%%%%%%%%%%%%%%%%%%%%%%%%%%%%%%
\begin{abstract}
We compute fragmentation functions for a quark to fragment to a
quarkonium through an $S$-wave spin-triplet heavy quark-antiquark pair.
We consider both color-singlet and color-octet heavy quark-antiquark
($Q\bar Q$) pairs. We give results for the case in which the fragmenting
quark and the quark that is a constituent of the quarkonium have
different flavors and for the case in which these quarks have the same
flavors. Our results for the sum over all spin polarizations of the
$Q\bar Q$ pairs confirm previous results. Our results for longitudinally
polarized $Q\bar Q$ pairs agree with previous calculations for the same 
flavor cases and correct an error in a previous calculation for the 
different-flavor case.
\end{abstract}
%12.38.Bx Perturbative calculations
%12.38.Cy Summation of perturbation theory
%12.39.St Factorization
%13.87.Fh Fragmentation into hadrons
%14.40.Pq Heavy quarkonia
\maketitle

\section{Introduction}

Quarkonium production at large transverse quarkonium momentum $p_T$
proceeds at leading power (LP) in $p_T$ through processes in which a
high-energy collision produces a single parton, which subsequently
fragments into a quarkonium \cite{Collins:1981uw}. Fragmentation
functions for a parton to fragment into a quarkonium play a central role
in calculations of such processes. In this paper, we compute the
fragmentation functions for a quark $q$ to fragment into a quarkonium
through heavy quark-antiquark ($Q\bar Q$) pair channels in which the
$Q\bar Q$ pair is in a spin-triplet $S$-wave state and a
color-singlet or a color-octet state. We calculate fragmentation
functions for the case in which the flavors of $q$ and $Q$ are the same,
as well as for the case in which the flavors are different. We carry
out these calculations at the leading nontrivial order in the strong
coupling $\alpha_s$ and at order $v^0$, where $v$ is the relative
velocity of the $Q$ and the $\bar Q$ in the quarkonium rest frame.

Previous calculations have given the fragmentation functions for a
quark to fragment into an $S$-wave, spin-triplet $Q\bar Q$ pair for the
case in which a sum over the $Q\bar Q$ spin polarizations has been
taken. The case in which the initial quark and final quark have
different flavors and are in a color-octet state is discussed in
Refs.~\cite{Ma:1995vi,Ma:2013yla}. The cases in which the initial quark
and final quark have the same flavor and are in a color-octet or
a color-singlet state is discussed in Ref.~\cite{Ma:2013yla}. Our
calculations confirm all of these results. We also verify a previous
calculation of the fragmentation function for a quark to fragment into
an $S$-wave, spin-triplet, color-singlet $Q\bar Q$ pair in which a sum
over the $Q\bar Q$ spin polarizations is taken \cite{Braaten:1993mp}.

We have extended all of these spin-summed calculations to the cases in
which the $Q\bar Q$ pair is in a longitudinally polarized state. After
our calculation was completed, we learned that these
longitudinal-polarization fragmentation functions had been calculated in
Ref.~\cite{hong-zhang}. Our calculation agrees with the results in
Ref.~\cite{hong-zhang} for the color-octet and color-singlet same-flavor
cases and corrects an error in Ref.~\cite{hong-zhang} for the
color-octet different-flavor case.

The remainder of this paper is organized as follows. In
Sec.~\ref{sec:notation}, we introduce our notation and the kinematics
that we use in the calculation and present projectors for the $Q\bar
Q$ spin and color. In Sec.~\ref{sec:collins-soper}, we present the
Collins-Soper fragmentation function for an initial quark
\cite{Collins:1981uw} and give the Feynman rules for its computation.
Sections \ref{sec:unequal-flavors} and \ref{sec:equal-flavors} contain,
respectively, the calculations of the color-octet fragmentation
functions for the case in which the initial and final quarks have
different flavors and the case in which the initial and final quarks
have the same flavor. In Sec.~\ref{sec:color-singlet}, we present the
fragmentation functions for the color-singlet case. Section
\ref{sec:summary} contains a summary and discussion of our results.

%=======================================================
\section{Notation, kinematics, and projectors \label{sec:notation}}
%=======================================================

In this paper, we use the following light-cone coordinates for a
four-vector $V$ in the $d=4-2\epsilon$ space-time dimensions:
\begin{subequations}
\begin{eqnarray}
V&=&(V^+,V^-,\bm{V}_\perp),\\
V^+&=&(V^0+V^{d-1})/\sqrt{2},\\
V^-&=&(V^0-V^{d-1})/\sqrt{2}, 
\end{eqnarray}
\end{subequations}
where we call $V^{d-1}$ the longitudinal component of the 
$(d-1)$-dimensional 
spatial vector $\bm{V}$, and $\bm{V}_\perp$ is the
$(d-2)$-dimensional component of $\bm{V}$ that is transverse to
$V^{d-1}$. In this coordinate system, the scalar product of  
two four-vectors $V$ and $W$ is given by $V\cdot
W=V^+W^-+V^-W^+-\bm{V}_\perp\cdot \bm{W}_\perp$.

At the leading nontrivial order in $\alpha_s$, a quark fragments into the
$Q\bar Q$ pair that forms the quarkonium plus an additional final-state
quark. We denote the momentum and mass 
of the fragmenting quark by $k$ and $m_q$, respectively. We denote the
momentum of the final-state quark by $k_1$, and we denote the momentum
of the $Q\bar Q$ pair by $P$.  We work at order $v^0$, and so we take
the $Q$ and the $\bar Q$ to have identical momenta $p = P/2$, with $p^2
= m_Q^2$, where $m_Q$ is the mass of the $Q$ or $\bar Q$. The mass of
the $Q\bar Q$ state is then given by $M=2m_Q$. 

We work in the frame in which the transverse momentum of the
$Q \bar Q$ pair vanishes. In this frame, the momenta are
%---------------
\begin{subequations}
\begin{eqnarray}
%---------------
k &=& 
\left( k^+, k^- = \frac{k^2+(P_\perp/z)^2}{2 k^+}, - \frac{\bm{P}_\perp}{z}
\right),
\\
P &=& 
\left( z k^+, \frac{M^2}{2 z k^+}, \bm{0}_\perp \right),
\\
k_1 &=& 
\left( z_1 k^+, \frac{m_q^2+k_{1\perp}^2}{2 z_1 k^+}, \bm{k}_{1 \perp} 
=-\frac{\bm{P}_\perp}{z}\right), 
%---------------
\end{eqnarray}
\end{subequations}
%---------------
where $z$ and $z_1\equiv 1-z$ are the longitudinal momentum fractions 
of the $Q\bar{Q}$ pair and the final-state quark, respectively:
%---------------
\begin{subequations}
\begin{eqnarray}
%---------------
z &=& \frac{P^+}{k^+},
\\
z_1 &=& \frac{k_1^+}{k^+}. 
%---------------
\end{eqnarray}
\end{subequations}
%---------------
We note that
%---------------
\begin{equation}
%---------------
k_1 \cdot P = P^+ k_1^- + P^- k_1^+ 
= \frac{z^2 (m_q^2+k_{1\perp}^2) + z_1^2 M^2}{2 z z_1}.
%---------------
\end{equation}
%---------------

We wish to project the $Q\bar Q$ pair onto spin-triplet states. The 
required spin-triplet projectors in order $v^0$ are 
\cite{Barbieri:1975am,Barbieri:1976fp,Chang:1979nn,Guberina:1980dc,Berger:1980ni,Bodwin:2002hg}
\begin{subequations}
\begin{eqnarray}
\Pi_3(p,p,\lambda)&=&- \frac{1}{2 \sqrt{2} m_Q} \epsilon\sl^* (\lambda) 
(p\sl + m_Q),\\
\bar\Pi_3(p,p,\lambda)&=&\gamma^0 {\Pi}_3^\dagger(p,p,\lambda)\gamma^0=
\frac{1}{2\sqrt{2}m_Q}\epsilon\sl (\lambda) 
(p\sl - m_Q),
\end{eqnarray}
where $\epsilon(\lambda)$ is the polarization vector for the spin state 
$\lambda$. 
\end{subequations}
These projectors correspond to nonrelativistic normalization of the
heavy-quark spinors.

The absolute squares of $\epsilon(\lambda)$ for various polarization 
states can be written in covariant forms. The result for the 
sum over all $\lambda$ is 
\begin{subequations}
\label{polarization-sums}
%---------------
\begin{equation}
%---------------
I_{\mu \nu} \equiv 
\sum_{\lambda=0, \pm1} \epsilon^*_\mu (\lambda) \epsilon_\nu (\lambda)
= 
-g_{\mu \nu} + \frac{P_\mu P_\nu}{P^2}.
%---------------
\end{equation}
%---------------
The result for the sum over transverse polarizations is~\cite{Cho:1994gb} 
%---------------
\begin{equation}
%---------------
I_{\mu \nu}^T \equiv
\sum_{\lambda=\pm1} \epsilon^*_\mu (\lambda) \epsilon_\nu (\lambda)
= 
-g_{\mu \nu} + 
\frac{P_\mu n_\nu + P_\nu n_\mu}{n \cdot P} 
- \frac{P^2}{(n \cdot P)^2} n_\mu n_\nu,
%---------------
\end{equation}
%---------------
where 
\begin{equation}
n=(0,1,\bm{0}_{\perp}).
\end{equation} 
Then, for the longitudinal polarization, we have
\begin{equation}
%---------------
I_{\mu \nu}^L \equiv \epsilon^*_\mu (0) \epsilon_\nu (0)=
I_{\mu \nu}-I_{\mu \nu}^T.
\end{equation}
\end{subequations}

The color-singlet and color-octet projection operators for the $Q\bar Q$ 
pair are
%---------------
\begin{subequations}
\begin{eqnarray}
%---------------
\Lambda_1&=&\frac{\bm{1}}{\sqrt{N_c}},\\
\Lambda_8^a &=&\sqrt{2}\, T^a,
%---------------
\end{eqnarray}
\end{subequations}
%---------------
where $\bm{1}$ is a unit $\textrm{SU}(N_c)$-color matrix,
$T^a$ is a generator of the fundamental representation of 
$\textrm{SU}(N_c)$,
$a\in \{1,\,2,\,\ldots,\,N_c^2-1\}$, and $N_c=3$.

%=======================================================
\section{Collins-Soper Definition of Fragmentation Function 
\label{sec:collins-soper}}
%=======================================================

Collins and Soper have given the following gauge-invariant definition of 
the quark fragmentation function in $d$ dimensions \cite{Collins:1981uw}:
%---------------
\begin{equation}
%---------------
D_{q\to H} (z) = 
\frac{z^{d-3}}{N_c \times 4 \times 2 \pi} 
\int_{-\infty}^{+\infty}
dx^- e^{-i P^+ x^-/z} 
{\rm tr} \left[n\sl \langle 0 | \Psi (0) 
{\cal E}^\dag (0) {\cal P}_{H (P,\lambda)} 
{\cal E} (x^-) \bar \Psi (x) |0 \rangle \right],
\label{collins-soper-frag}
%---------------
\end{equation}
%---------------
where $\Psi(x)$ is
the field of the initial quark and ${\cal E} (x^-)$ is the gauge link 
(eikonal line) 
%---------------
\begin{equation} 
%---------------
{\cal E} (x^-)={\cal P}\exp\biggl[+ig_s\int_{x^-}^\infty
dz^-\, A^+(0^+,z^-,\bm{0}_\perp)\biggr].
%---------------
\end{equation}
%---------------
Here, $\mathcal{P}$ indicates path ordering, and $g_s=\sqrt{4\pi
\alpha_s}$ is the QCD coupling constant. The spinor field $\Psi$ is an
$\textrm{SU}(N_c)$-color column vector in the fundamental
representation, and the gluon field $A^\mu=A^\mu_a T^a$ is an
$\textrm{SU}(N_c)$-color matrix in the fundamental representation. The
trace is over the color and Dirac indices. The factors $N_c$ and $4$ in
the denominator of Eq.~(\ref{collins-soper-frag}) arise from the
average over the color and Dirac indices of the initial-state quark,
respectively. ${\cal P}_{H (P,\lambda)}$ is a projector onto states that
include a hadron $H$ with momentum $P$ and polarization $\lambda$:
\begin{equation}
{\cal P}_{H (P,\lambda)}=\sum_X|H(P,\lambda)+X\rangle
\langle H(P,\lambda)+X|,
\label{frag-projector}
\end{equation}
where the summation is over all possible degrees of freedom.

The Feynman rules for the fragmentation function are the standard ones
for QCD, with the following exceptions.  First, there is an overall
factor
%---------------
\begin{equation}
%---------------
C_{\rm qfrag} = \frac{z^{1-2\epsilon}}{8\pi N_c},
%---------------
\label{norm-factor}
\end{equation}
%---------------
which arises from the definition of the fragmentation function. Second,
there are additional Feynman rules for the eikonal lines. We state the
rules for the part of the Feynman diagram that lies to the left of the
final-state cut. The rules for the part of the diagram that lies to the
right of the final-state cut can be obtained by complex conjugation.
Each eikonal-line propagator that carries momentum $\ell$, flowing
from the cut side to the operator side, contributes a factor
$i\delta_{ij}/(\ell\cdot n +i\varepsilon)$, where $i$ and $j$ are color
indices. Each eikonal-line--gluon vertex
contributes a factor $i g_s n_\mu T_{ij}^a$, where $\mu$ is the
four-vector index of the gluon. The final-state cut in an eikonal line
carrying momentum $\ell$ contributes a factor $2\pi\delta(\ell\cdot n)$.

In general, the final-state phase space for a fragmentation function 
for $n$ unobserved particles in the final state is given by
%---------------
\begin{equation}
%---------------
d\Phi_n=\frac{4\pi M}{S}\delta\biggl(k^+-P^+ -\sum_{i=1}^n k_i^+\biggr)
\left( \frac{\mu^2}{4 \pi} e^{\gamma_{\rm E}} \right)^{n\epsilon}
\prod_{i=1}^n \theta(k_i^+)
\frac{dk_i^+ d^{2-2\epsilon}\bm{k}_{i\perp}}
{2k_i^+(2\pi)^{3-2\epsilon}},
%---------------
\end{equation}
%---------------
where $S$ is the statistical factor for identical final-state particles,
$M$ and $P$ are the mass and momentum of the observed particle,
respectively, and $k_i$ is the momentum of the $i$th unobserved
particle. We have included a factor $2M$ in the phase space in order to
compensate for the fact that we use nonrelativistic normalization for
the heavy-quark spinors. We associate the standard
modified-minimal-subtraction ($\overline{\rm MS}$) scale factor
$[\mu^2e^{\gamma_{\rm E}}/(4\pi)]^\epsilon$
with each dimensionally regulated integration in $d=4-2\epsilon$
space-time dimensions. Here, $\mu$ is the dimensional-regularization scale 
and $\gamma_{\rm E}$ is the Euler-Mascheroni constant.

For our specific kinematics, with one unobserved particle in the final 
state, the phase space reduces to 
%---------------
\begin{eqnarray}
%---------------                                            
d \Phi &=&                                                        
4 \pi M  \delta (k^+ - P^+-k_1^+)
\left( \frac{\mu^2}{4 \pi} e^{\gamma_{\rm E}} \right)^{\epsilon}
 \theta(k^+)\frac{dk_1^+}{4 \pi k_1^+}
\frac{d^{d-2} \bm{k}_{1 \perp}}{(2 \pi)^{d-2}}
\nonumber \\
&=&
\frac{4\pi M}{k^+} \delta (1-z-z_1)
\left( \frac{\mu^2}{4 \pi} e^{\gamma_{\rm E}} \right)^{\epsilon}
\theta(z_1) \frac{dz_1}{4 \pi z_1}
\frac{d^{2-2\epsilon} \bm{k}_{1 \perp}}{(2 \pi)^{2-2\epsilon}}.
\label{2p-PS}
%---------------
\end{eqnarray}
%---------------

%=======================================================
\section{Color-octet fragmentation: different-flavor case
\label{sec:unequal-flavors}}

\begin{figure}
\epsfig{file=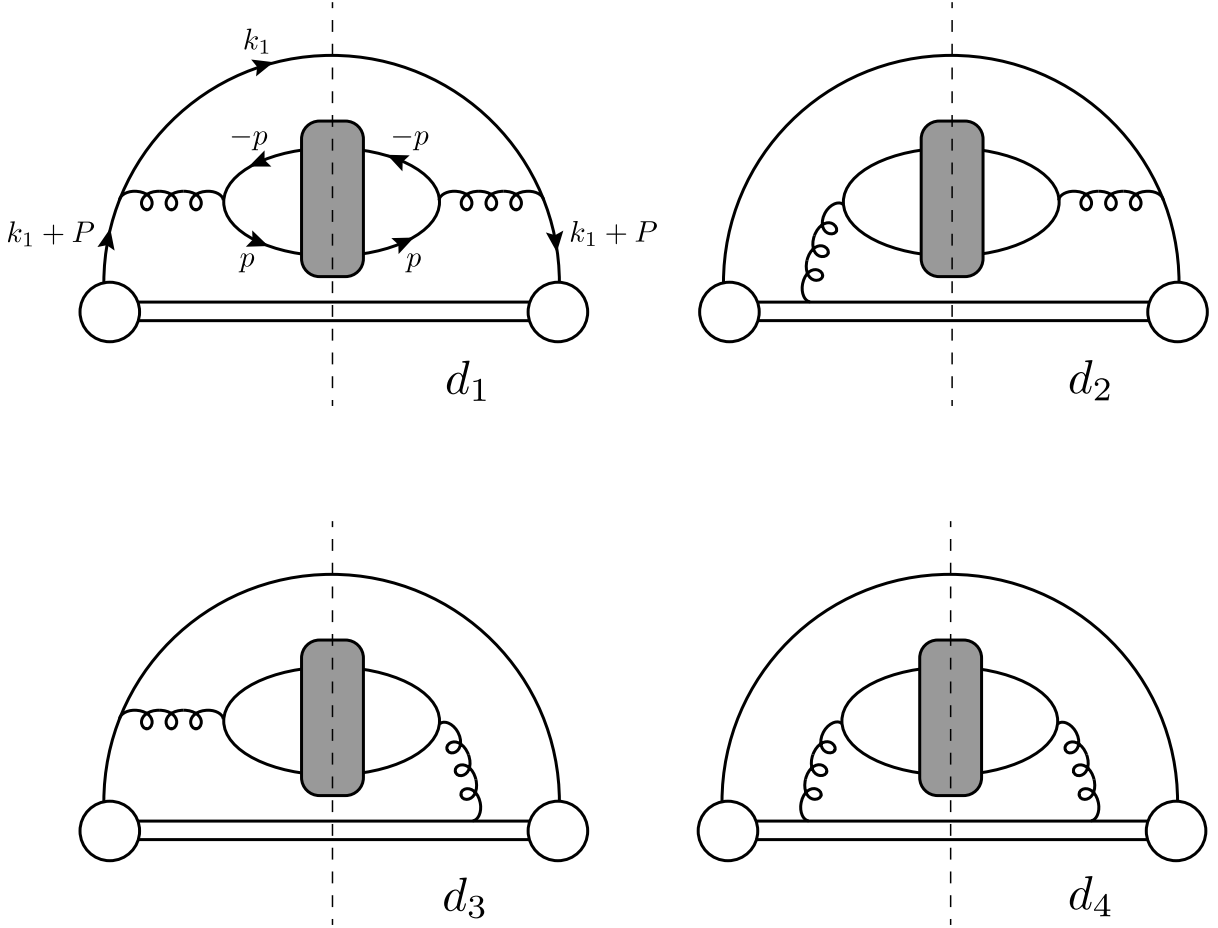,width=16cm}
\caption{Feynman diagrams for quark fragmentation into a color-octet 
$Q\bar Q$ pair for the case in which the initial             
quark $q$ and the quark $Q$ that is a constituent of the quarkonium have
different flavors. The diagram labels $d_i$ correspond to the quantities that 
appear in Eq.~(\ref{unequal-frag}). \label{fig:unequal}}
\end{figure}

In this section, we compute the fragmentation function for quark fragmentation
into a color-octet $Q\bar Q$ pair for the case in which the initial
quark $q$ and the quark $Q$ that is a constituent of the quarkonium have
different flavors. The Feynman diagrams for this calculation are shown 
in Fig.~\ref{fig:unequal}. In this calculation, and throughout the 
remainder of this paper, we work in the Feynman gauge.

In each of the contributions from the diagrams in
Fig.~\ref{fig:unequal}, there is a common factor that arises from the
annihilation of a virtual gluon into a color-octet spin-triplet $S$-wave
pair $Q \bar{Q}({}^3S^{[8]}_1$). The contribution to this factor from
the left side of the final-state cut can be written as
%---------------
\begin{equation}
%---------------
J_\mu^{ab}(\lambda) = \frac{-i g_{\mu \nu}}{P^2+i \varepsilon} 
{\rm tr} [ (-i g_s \gamma^\nu T^a) \Pi_3 (p,p,\lambda) \Lambda_8^b].
%---------------
\end{equation}
%---------------
A straightforward calculation gives
%---------------
\begin{equation}
%---------------
J_\mu^{ab} (\lambda) =   
\frac{g_s}{M^2+i \varepsilon}
\delta^{ab} \epsilon^*_\mu (\lambda).
%---------------
\end{equation}
%---------------
Multiplying by the complex conjugate and summing over the final-state color
index, we obtain 
%---------------
\begin{equation}
%---------------
X_{\mu \nu}^{ab} (\lambda) = J_\mu^{ac} (\lambda) J_\nu^{bc}{}^* (\lambda)
=\frac{g_s^2}{M^4}
\delta^{ab} \epsilon^*_\mu (\lambda)
\epsilon_\nu (\lambda). 
%---------------
\end{equation}
%---------------

Then, the diagrams of Fig.~\ref{fig:unequal} give the following 
contributions to the fragmentation function:
%---------------
\begin{subequations}
\label{unequal-frag}
\begin{eqnarray}
%---------------
d_1(z) &=& C_{\rm qfrag} 
{\rm tr} \bigg[ 
(k \sl_1+m_q) (-i g_s \gamma^\mu T^a) 
\frac{i}{k \sl_1+P\sl-m_q+i \varepsilon} n \sl
\nonumber \\ && \hspace{20ex} \times 
\frac{-i}{k \sl_1+P\sl-m_q-i \varepsilon} 
(+i g_s \gamma^\nu T^b) 
\bigg] X_{\mu \nu}^{ab} d \Phi,
\hspace{5ex}
\\
d_2(z) &=& C_{\rm qfrag} 
{\rm tr} \bigg[ 
(k \sl_1 +m_q)
\frac{i}{(k-k_1) \cdot n+i \varepsilon}
(i g_s n^\mu T^a) 
n \sl
\nonumber \\ && \hspace{20ex} \times 
\frac{-i}{k \sl_1+P\sl-m_q-i \varepsilon} 
(+i g_s \gamma^\nu T^b) 
\bigg] X_{\mu \nu}^{ab} d \Phi,
\\
d_3(z) &=& C_{\rm qfrag} 
{\rm tr} \bigg[ 
(k \sl_1+m_q) (-i g_s \gamma^\mu T^a) 
\frac{i}{k \sl_1+P\sl-m_q+i \varepsilon} n \sl
\nonumber \\ && \hspace{20ex} \times 
(-i g_s n^\nu T^b) 
\frac{-i}{(k-k_1) \cdot n-i \varepsilon}
\bigg] X_{\mu \nu}^{ab} d \Phi,
\\
d_4(z) &=& C_{\rm qfrag} 
{\rm tr} \bigg[ 
(k \sl_1 +m_q)
\frac{i}{(k-k_1) \cdot n+i \varepsilon} 
(+i g_s n^\mu T^a) 
n \sl
\nonumber \\ && \hspace{20ex} \times 
(-i g_s n^\nu T^b) 
\frac{-i}{(k-k_1) \cdot n-i \varepsilon}
\bigg] X_{\mu \nu}^{ab} d \Phi.
%---------------
\end{eqnarray}
\end{subequations}
%---------------
In each contribution in Eq.~(\ref{unequal-frag}), the overall color
factor, including the color factor in $X_{\mu\nu}^{ab}$, is
\begin{equation}
\frac{1}{2}\delta^{ab}{\rm tr}(T^aT^b)=\frac{C_F N_c}{2},
\end{equation}
where $C_F=(N_c^2-1)/(2N_c)$.
The dependence on $\bm{k}_{1\perp}$ in these expressions comes from the
quark-propagator denominators, which contribute factors 
$(\bm{k}_{1\perp}^2 +m_q^2 + \tfrac{1-z}{z^2} M^2)^{-1}$.
Thus, 
the
integrals of these expressions over $\bm{k}_{1\perp}$ can be written in
terms of 
the scalar integrals 
$J_n (m_q^2 + \tfrac{1-z}{z^2} M^2)$, where
%---------------
\begin{eqnarray}
%---------------
J_n (s) &\equiv& 
\left( \frac{\mu^2}{4 \pi} e^{\gamma_{\rm E}} \right)^{\epsilon}
\int \frac{d^{2-2\epsilon} \bm{k}_{1\perp}}{(2 \pi)^{2-2\epsilon}}
\frac{1}{(\bm{k}_{1\perp}^2 + s)^{n}}
\nonumber\\
&=& 
\frac{\left( \mu^2 e^{\gamma_{\rm E}} \right)^{\epsilon}}{4 \pi} 
\frac{\Gamma(n-1+\epsilon)}{\Gamma(n)} 
s^{1-n-\epsilon}. 
%---------------
\label{Jn}
\end{eqnarray}
The only integral that diverges as $\epsilon\to 0$ is $J_1(s)$:
\begin{equation}
J_1(s)=
\frac{
\left( \mu^2 e^{\gamma_{\rm E}}/s
\right)^{\epsilon}}{4 \pi} 
\frac{\Gamma(1+\epsilon)}{\epsilon}.
\end{equation}
For $n\geq 2$, $J_n (s)$, we can write
%---------------
\begin{eqnarray}
%---------------
J_n (s) &=& 
\frac{
s^{1-n}
}{4 \pi (n-1)} 
+O(\epsilon)\quad ( n\geq 2). 
%---------------
\end{eqnarray}
%---------------

Summing over the four contributions in Eq.~(\ref{unequal-frag}),  
using the expressions for the absolute squares of the polarizations in 
Eq.~(\ref{polarization-sums}), multiplying by the factor in 
Eq.~(\ref{norm-factor}), and carrying out the phase-space 
integration in Eq.~(\ref{2p-PS}), we obtain the following results for the 
fragmentation functions:
%---------------
\begin{subequations}
\begin{eqnarray}
%---------------
&& 
\sum_\lambda D_{q\to Q \bar Q({}^3S_1^{[8]})(\lambda)}^{(1-4)}(z)
= 
\frac{g_s^4 C_F}{\pi M^3 z^{1+2 \epsilon}} 
\{ [1+ (1-z)^2-\epsilon z^2] J_1 (m_q^2 + \tfrac{1-z}{z^2} M^2)
\nonumber \\ && \hspace{32ex} - [(1-\epsilon) M^2 +2 m_q^2] (1-z) J_2 
(m_q^2 + \tfrac{1-z}{z^2} M^2)\},\phantom{xxx}
\\
&& 
D_{q\to Q \bar Q({}^3S_1^{[8]})(\lambda =0)}^{(1-4)}(z)
= 
\frac{2 g_s^4 C_F}{\pi M} 
\frac{(1-z)^2}{z^{3+2 \epsilon}} J_2(m_q^2 + \tfrac{1-z}{z^2} M^2). 
%---------------
\end{eqnarray}
\end{subequations}
%---------------
Here, we retain the full $\epsilon$ dependence, as it may be useful for 
calculations of fragmentation functions at higher orders in $\alpha_{s}$.

The expression for $\sum_\lambda D_{q\to Q \bar Q({}^3S_1^{[8]})(\lambda)}$ 
contains a pole in $\epsilon$. We renormalize this expression using the  
$\overline{\rm MS}$ procedure \cite{Collins:1981uw}:
%---------------
\begin{equation}
%---------------
D^{{\overline{\rm MS}}}_{q\to A} (z, \mu) = 
D_{q\to A} (z, \mu) - \frac{1}{\epsilon}
 \frac{\alpha_s}{2 \pi}
 \int_{z}^{1} \frac{dy}{y}\, 
P_{gq}(z/y)\,
D_{g\to A} (y), 
%---------------
\end{equation}
%---------------
where the $D_{q\to A}(z)$ and the $D_{g\to A}(z)$ are the bare quark and gluon
fragmentation functions and 
$P_{gq}(z)$ is the Dokshitzer-Gribov-Lipatov-Altarelli-Parisi 
splitting function, which is given at lowest order in $\alpha_s$ by
%---------------
\begin{equation}
%---------------
P_{gq} (z) = C_F \frac{1+(1-z)^2}{z}. 
%---------------
\end{equation}
%---------------
The bare gluon fragmentation functions for the unpolarized and
longitudinally polarized states are given at leading order in $\alpha_s$
by~\cite{Bodwin:2012xc}
%---------------
\begin{subequations}
\label{bare-gluon-frag}
\begin{eqnarray}
%---------------
&&
\sum_\lambda D_{g\to Q \bar Q({}^3S_1^{[8]})(\lambda)} (z) = 
\frac{\pi \alpha_s}{m_Q^3} \delta(1-z), 
\\
&& 
D_{g\to Q \bar Q({}^3S_1^{[8]})(\lambda = 0 )}(z) = 0.
%---------------
\end{eqnarray}
\end{subequations}
%---------------
At this order in $\alpha_s$, the bare gluon fragmentation functions do
not depend on $\epsilon$.\footnote{ The $\epsilon$ dependence in
Eq.~(\ref{bare-gluon-frag}) is different from that in the corresponding
expression in Ref.~\cite{Bodwin:2012xc}. In Ref.~\cite{Bodwin:2012xc}, a
factor $[\mu^2 e^{\gamma_{\rm E}}/(4 \pi)]^\epsilon$ was associated with
each factor $g_s^2$. In the present paper, we associate a factor
$[\mu^2 e^{\gamma_{\rm E}}/(4 \pi)]^\epsilon$ with each dimensionally
regulated integral. This difference between these conventions does not
affect the finite result. } Carrying out the renormalization and 
dropping contributions of order $\epsilon$ and higher, we obtain
%---------------
\begin{subequations}
\label{QQbar-unequal}
\begin{eqnarray}
%---------------
&& 
\sum_\lambda D_{q\to Q \bar Q({}^3S_1^{[8]})(\lambda)}^{\overline{\rm MS}}
(z, \mu)
= 
\frac{\alpha_s^2 C_F}{2 m_Q^3} 
\bigg\{ 
\frac{z^2-2z+2}{z} 
\bigg[ \log \frac{\mu^2}{4 m_Q^2} - \log (1-z+r z^2) \bigg] 
\nonumber \\ 
&& \hspace{35ex} 
- z 
- 
\frac{ z (1-z) (1+2 r) }{1-z + r z^2} \bigg\}
\frac{
\langle O^{Q \bar Q({}^3S_1^{[8]})}({}^3S_1^{[8]}) \rangle
}
{3 (N_c^2-1)}
, 
\\
&& 
D_{q\to Q \bar Q({}^3S_1^{[8]})(\lambda =0)}(z)
= 
\frac{\alpha_s^2 C_F}{2 m_Q^3} 
\frac{2(1-z)}{z} 
\frac{1-z}{1-z + r z^2} 
\frac{\langle O^{Q \bar Q({}^3S_1^{[8]})}({}^3S_1^{[8]}) \rangle}
{3 (N_c^2-1)}, 
%---------------
\end{eqnarray}
\end{subequations}
%---------------
where $ r \equiv m_q^2/M^2 = m_q^2/(2 m_Q)^2$. 
In Eq.~(\ref{QQbar-unequal}), we have written the perturbative
fragmentation function in the factorized form of a short-distance
coefficient times a nonrelativistic-QCD (NRQCD) long-distance matrix element 
(LDME) \cite{Bodwin:1994jh} by making use of the fact\footnote{Our convention 
for the LDME is to sum over
all spin states of the quarkonium or $Q\bar Q$ pair, even in the case of
fragmentation into a single (longitudinally polarized) spin state. In
that case, we use the fact that the LDMEs for different spin
states are identical, up to corrections of relative order $v^2$.
Note that our LDMEs for the longitudinally polarized case are
larger than those in Ref.~\cite{hong-zhang} by a factor of three.}
that, at order
$\alpha_s^{0}$,
\begin{equation}
\langle O^{Q \bar Q({}^3S_1^{[8]})}({}^3S_1^{[8]}) \rangle = 
(N_c^2-1)(d-1).
\end{equation}
The NRQCD LDME  $\langle O^{Q \bar Q({}^3S_1^{[8]})}({}^3S_1^{[8]})
\rangle$ is defined by \begin{equation} \langle O^{Q \bar
Q({}^3S_1^{[8]})}({}^3S_1^{[8]}) \rangle=\langle 0|
\chi^\dagger\sigma^iT^a\psi \sum_{\lambda}{\cal P}_{Q \bar
Q({}^3S_1^{[8]})(P,\lambda)} \psi^\dagger\sigma^iT^a\chi|0\rangle.
\end{equation} Here, $\psi$ is the two-component (Pauli) spinor field
operator that annihilates a heavy quark and $\chi^\dagger$ is the
two-component (Pauli) spinor field operator that annihilates a heavy
antiquark. The projection operator ${\cal P}_{Q \bar
Q({}^3S_1^{[8]})(P,\lambda)}$ is the free $Q\bar{Q}$ analogue of ${\cal
P}_{H(P,\lambda)}$, except that, because we are considering an NRQCD
LDME, the intermediate state can contain only light degrees of freedom
in addition to the explicit $Q\bar Q$ pair.

We can now obtain the quarkonium fragmentation functions for the
unpolarized and the longitudinally polarized states by replacing the
free $Q\bar Q$ LDMEs in Eq.~(\ref{QQbar-unequal}) with the
quarkonium LDMEs:
%---------------
\begin{subequations}
\label{ms-bar-frag-unequal}
\begin{eqnarray}
%---------------
\label{ms-bar-frag-unequal-sum}
&& 
\sum_\lambda D_{q\to H(\lambda)}^{\overline{\rm MS}}
(z, \mu)
= 
\frac{\alpha_s^2 C_F}{2 m_Q^3} 
\bigg\{ 
\frac{z^2-2z+2}{z} 
\bigg[ \log \frac{\mu^2}{4 m_Q^2} - \log (1-z+r z^2) \bigg] 
\nonumber \\ 
&& \hspace{35ex} 
- z 
- 
\frac{ z (1-z) (1+2 r) }{1-z + r z^2} \bigg\}
\frac{\langle O^H({}^3S_1^{[8]}) \rangle}
{3 (N_c^2-1)}
, 
\\
\label{ms-bar-frag-unequal-long}
&& 
D_{q\to H(\lambda=0)}(z) 
= 
\frac{\alpha_s^2 C_F}{2 m_Q^3} 
\frac{2(1-z)}{z} 
\frac{1-z}{1-z + r z^2} 
\frac{\langle O^H({}^3S_1^{[8]}) \rangle}
{3 (N_c^2-1)},
%---------------
\end{eqnarray}
\end{subequations}
%---------------
where the NRQCD LDME $\langle 
O^H({}^3S_1^{[8]}) \rangle$ is defined by
\begin{equation}
\langle O^H({}^3S_1^{[8]}) \rangle=\langle 0|
\chi^\dagger\sigma^iT^a\psi\sum_{\lambda}{\cal P}_{H(P,\lambda)}
\psi^\dagger\sigma^iT^a\chi|0\rangle.
\end{equation}
Identical expressions hold for the case of an initial antiquark.

The result for the polarization-summed fragmentation function in
Eq.~(\ref{ms-bar-frag-unequal-sum}) confirms the result in Eq.~(4.2) of
Ref.~\cite{Ma:1995vi} and the result in Eq.~(C37) of
Ref.~\cite{Ma:2013yla}. The result for the
longitudinal-polarization fragmentation function in
Eq.~(\ref{ms-bar-frag-unequal-long}) disagrees with the result in
Eq.~(C.10) of Ref.~\cite{hong-zhang}.\footnote{We find that a
denominator factor in Eq.~(C.10) of Ref.~\cite{hong-zhang} should be
$\eta z^2-4 z+4$, rather than $\eta^2 z^2-4 z+4$, where
$\eta=m_q^2/m_Q^2=4r$. We also find that the result in Eq.~(C.10) of
Ref.~\cite{hong-zhang} should be multiplied by an overall factor of three.
Here, we have taken into account the fact that the LDME in
Eq.~(\ref{ms-bar-frag-unequal-long}) is a factor of three larger than the
corresponding LDME in Ref.~\cite{hong-zhang}.} The author of
Ref.~\cite{hong-zhang} has confirmed that the result in
Eq.~(\ref{ms-bar-frag-unequal-long}) is correct.

In the case of light initial quarks, it is useful to take the limit 
$m_q\to 0$, which gives
%---------------
\begin{subequations}
\label{unequal-zero-mass}
\begin{eqnarray}
%---------------
\sum_{\lambda} D_{q\to H(\lambda)}^{\overline{\rm MS}}(z, \mu)
&=& \frac{\alpha_s^2 C_F}{2 m_Q^3} 
\left[ \frac{z^2-2z+2}{z} 
\log \frac{\mu^2}{4 m_Q^2 (1-z)} - 2 z \right]
\frac{\langle O^H({}^3S_1^{[8]}) \rangle}{3 (N_c^2-1)},
\label{unequal-zero-mass-sum-pol}\\
D_{q\to H(\lambda=0)}(z)
&=& 
\frac{\alpha_s^2 C_F}{2 m_Q^3}
\frac{ 2 (1-z) }{z}
\frac{\langle O^H({}^3S_1^{[8]}) \rangle}{3 
(N_c^2-1)}.\label{unequal-zero-mass-long-pol}
%---------------
\end{eqnarray}
\end{subequations}
%---------------
We have compared our result in Eq.~(\ref{unequal-zero-mass-sum-pol})
with the result in Eq.~(20) of Ref.~\cite{Braaten:2001sz} for quark
fragmentation into lepton pairs, taking into account differences in the
color and phase-space factors, and have found that they are consistent
with each other.

%=======================================================
\section{Color-octet fragmentation: same-flavor case
\label{sec:equal-flavors}} 
%=======================================================

\begin{figure}
\epsfig{file=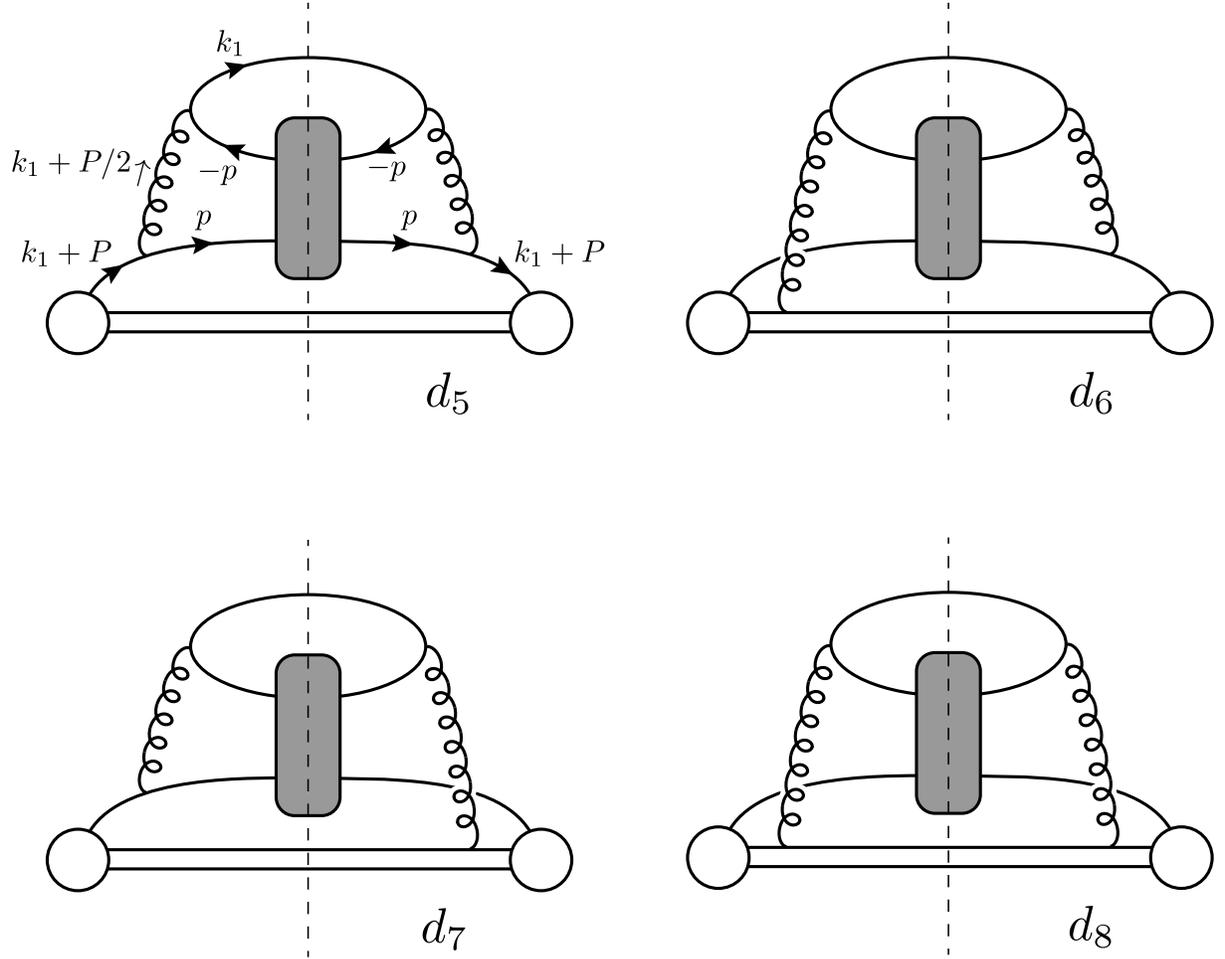,width=16cm}
\caption{Additional Feynman diagrams for quark fragmentation into a
color-octet $Q\bar Q$ pair for the case in which the initial quark $q$
and the quark $Q$ that is a constituent of the quarkonium have 
the same flavor. These diagrams differ from those in
Fig.~\ref{fig:unequal} in that the identical quarks have been
interchanged in the amplitudes on both the left and the right sides of the
final-state cut. The diagram labels $d_i$ correspond to the quantities that 
appear in Eq.~(\ref{exchange-frag}). \label{fig:exchange}}
\end{figure}

\begin{figure}
\epsfig{file=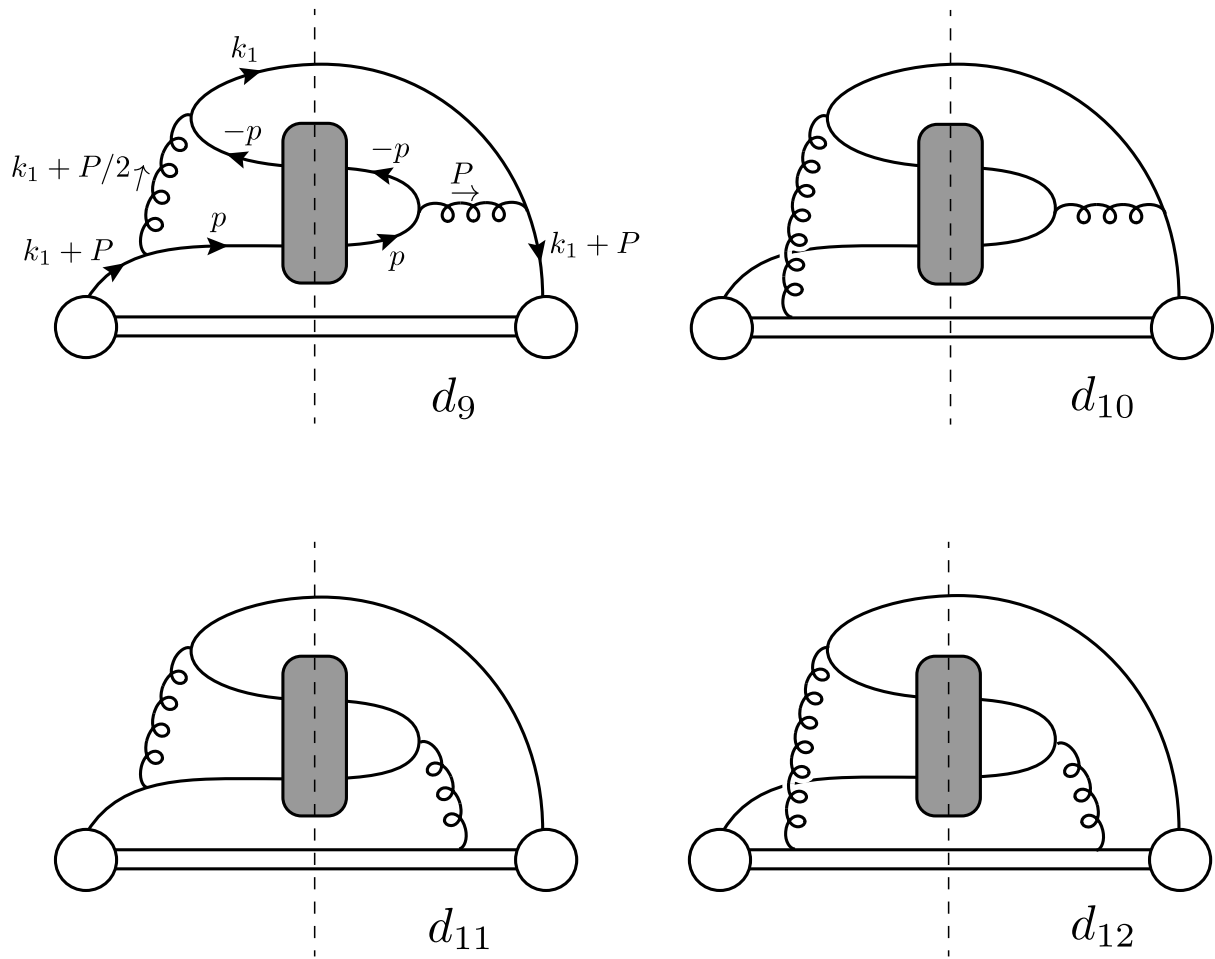,width=16cm}
\caption{Additional Feynman diagrams for quark fragmentation into a
color-octet $Q\bar Q$ pair for the case in which the initial quark $q$
and the quark $Q$ that is a constituent of the quarkonium have the same
flavor. These diagrams differ from those in Fig.~\ref{fig:unequal} in
that the identical quarks have been interchanged in an amplitude on only
one side of the final-state cut. The diagram labels $d_i$ correspond 
to the quantities that appear in Eq.~(\ref{interference-frag}).
\label{fig:interference}}
\end{figure}

Now let us consider the fragmentation function for the case in which the
initial quark $q$ and the quark $Q$ that is a constituent of the quarkonium
have the same flavor. In this case, there are contributions from the
diagrams that are shown in Fig.~\ref{fig:unequal}. These are given by
the expressions in Eq.~(\ref{unequal-frag}), but with $m_q$ set equal to
$m_Q$. In addition, there are contributions from the diagrams that are
shown in Fig.~\ref{fig:exchange} and in Fig.~\ref{fig:interference}. The
diagrams in Fig.~\ref{fig:exchange} differ from those in
Fig.~\ref{fig:unequal} in that the identical quarks have been
interchanged in the amplitudes on both the left and right sides of the
final-state cut. The diagrams in Fig.~\ref{fig:interference} differ from
those in Fig.~\ref{fig:unequal} in that the identical quarks have been
interchanged in an amplitude on only one side of the cut.

The contributions of the diagrams in Fig.~\ref{fig:exchange} are
%---------------
\begin{subequations}
\label{exchange-frag}
\begin{eqnarray}
%---------------
d_5 (z,\lambda) &=& 
C_{\rm qfrag}
{\rm tr} \bigg[
\Pi_3 (p,p,\lambda) \Lambda_8^c 
(-i g_s \gamma^\mu T^a) \frac{i}{k \sl_1 + P \sl-m_Q+i\varepsilon} 
n \sl 
\frac{-i}{k \sl_1 + P \sl-m_Q-i\varepsilon} 
\nonumber \\ && \hspace {15ex} \times 
(+i g_s \gamma^\nu T^b) 
\bar{\Pi}_3 (p,p,\lambda) \Lambda_8^c
(+i g_s \gamma_\nu T^b) (k \sl_1 + m_Q) 
(-i g_s \gamma_\mu T^a) 
\bigg]
\nonumber \\ && \hspace {15ex} \times 
\frac{-i}{(k_1+P/2)^2+i \varepsilon} 
\frac{+i}{(k_1+P/2)^2-i \varepsilon} 
d \Phi, 
\\
d_6 (z,\lambda) &=& 
C_{\rm qfrag}
{\rm tr} \bigg[
\Pi_3 (p,p,\lambda) \Lambda_8^c 
\frac{i}{(k-P/2)\cdot n+i\varepsilon} (i g_s n^\mu T^a) 
n \sl 
\frac{-i}{k \sl_1 + P \sl-m_Q-i\varepsilon} 
\nonumber \\ && \hspace {15ex} \times 
(+i g_s \gamma^\nu T^b) 
\bar{\Pi}_3 (p,p,\lambda) \Lambda_8^c
(+i g_s \gamma_\nu T^b) (k \sl_1 +m_Q) 
(-i g_s \gamma_\mu T^a) 
\bigg]
\nonumber \\ && \hspace {15ex} \times 
\frac{-i}{(k_1+P/2)^2+i \varepsilon} 
\frac{+i}{(k_1+P/2)^2-i \varepsilon} 
d \Phi, 
\\
d_7 (z,\lambda) &=& 
C_{\rm qfrag}
{\rm tr} \bigg[
\Pi_3 (p,p,\lambda) \Lambda_8^c 
(-i g_s \gamma^\mu T^a) \frac{i}{k \sl_1 + P \sl-m_Q+i\varepsilon} 
n \sl 
(-i g_s n^\nu T^b) 
\nonumber \\ && \hspace {15ex} \times 
\frac{-i}{(k-P/2) \cdot n-i\varepsilon} 
\bar{\Pi}_3 (p,p,\lambda) \Lambda_8^c
(+i g_s \gamma_\nu T^b) (k \sl_1 + m_Q) 
(-i g_s \gamma_\mu T^a) 
\bigg]
\nonumber \\ && \hspace {15ex} \times 
\frac{-i}{(k_1+P/2)^2+i \varepsilon} 
\frac{+i}{(k_1+P/2)^2-i \varepsilon} 
d \Phi, 
\\
d_8 (z,\lambda) &=& 
C_{\rm qfrag}
{\rm tr} \bigg[
\Pi_3 (p,p,\lambda) \Lambda_8^c 
\frac{i}{(k-P/2)\cdot n+i\varepsilon} (i g_s n^\mu T^a) 
n \sl 
(-i g_s n^\nu T^b) 
\nonumber \\ && \hspace {15ex} \times 
\frac{-i}{(k-P/2) \cdot n-i\varepsilon} 
\bar{\Pi}_3 (p,p,\lambda) \Lambda_8^c
(+i g_s \gamma_\nu T^b) (k \sl_1 + m_Q) 
(-i g_s \gamma_\mu T^a) 
\bigg]
\nonumber \\ && \hspace {15ex} \times 
\frac{-i}{(k_1+P/2)^2+i \varepsilon} 
\frac{+i}{(k_1+P/2)^2-i \varepsilon} 
d \Phi.
%---------------
\end{eqnarray}
\end{subequations}
%---------------
In each contribution in Eq.~(\ref{exchange-frag}), the overall color 
factor is
\begin{equation}
%---------------
{\rm tr} \left( \Lambda_8^c T^a T^b \Lambda_8^c T^b T^a \right)=
\frac{C_F}{2N_c}.
%---------------
\label{exchange-CF}
\end{equation}
%---------------

The contributions from the diagrams in Fig.~\ref{fig:interference} are 
%---------------
\begin{subequations}
\label{interference-frag}
\begin{eqnarray}
%---------------
d_9(z, \lambda) &=& 
-C_{\rm qfrag}
{\rm tr} \bigg[
\Pi_3 (p,p,\lambda) \Lambda_8^c 
(-i g_s \gamma^\mu T^a) \frac{i}{k \sl_1 + P \sl-m_Q+i\varepsilon} 
n \sl 
\nonumber \\ && \hspace {15ex} \times 
\frac{-i}{k \sl_1 + P \sl-m_Q-i\varepsilon} 
(+i g_s \gamma^\nu T^b) 
(k \sl_1 +m_Q) (-i g_s \gamma_\mu T^a) 
\bigg]
\times 
J_\nu^{bc}{}^*
\nonumber \\ && \hspace {15ex} \times 
\frac{-i}{(k_1+P/2)^2+i \varepsilon} 
d \Phi+{\rm c.c.}, 
\\
d_{10}(z, \lambda) &=& 
-
C_{\rm qfrag}
{\rm tr} \bigg[
\Pi_3 (p,p,\lambda) \Lambda_8^c 
\frac{i}{(k-P/2)\cdot n+i\varepsilon} (i g_s n^\mu T^a) 
n \sl 
\nonumber \\ && \hspace {15ex} \times 
\frac{-i}{k \sl_1 + P \sl-m_Q-i\varepsilon} 
(+i g_s \gamma^\nu T^b) 
(k \sl_1 +m_Q) (-i g_s \gamma_\mu T^a) 
\bigg]
\times 
J_\nu^{bc}{}^*
\nonumber \\ && \hspace {15ex} \times 
\frac{-i}{(k_1+P/2)^2+i \varepsilon} 
d \Phi+{\rm c.c.}, 
\\
d_{11}(z, \lambda) &=& 
-C_{\rm qfrag}
{\rm tr} \bigg[
\Pi_3 (p,p,\lambda) \Lambda_8^c 
(-i g_s \gamma^\mu T^a) \frac{i}{k \sl_1 + P \sl-m_Q+i\varepsilon} 
n \sl 
\nonumber \\ && \hspace {15ex} \times 
(-ig_sn^\nu T^b) \frac{-i}{(k-k_1) \cdot n -i \varepsilon} 
(k \sl_1 +m_Q) (-i g_s \gamma_\mu T^a) 
\bigg]
\times 
J_\nu^{bc}{}^*
\nonumber \\ && \hspace {15ex} \times 
\frac{-i}{(k_1+P/2)^2+i \varepsilon} 
d \Phi+{\rm c.c.}, 
\\
d_{12}(z, \lambda) &=& 
-
C_{\rm qfrag}
{\rm tr} \bigg[
\Pi_3 (p,p,\lambda) \Lambda_8^c 
\frac{i}{(k-P/2)\cdot n+i\varepsilon} (i g_s n^\mu T^a) 
n \sl 
\nonumber \\ && \hspace {15ex} \times 
(-ig_sn^\nu T^b) \frac{-i}{(k-k_1) \cdot n -i \varepsilon} 
(k \sl_1 +m_Q) (-i g_s \gamma_\mu T^a) 
\bigg]
\times 
J_\nu^{bc}{}^*
\nonumber \\ && \hspace {15ex} \times 
\frac{-i}{(k_1+P/2)^2+i \varepsilon} 
d \Phi+{\rm c.c.}, 
%---------------
\end{eqnarray}
\end{subequations}
%---------------
where c.c. stands for complex conjugate. 
In each contribution in
Eq.~(\ref{interference-frag}), the overall color factor, 
including the color factor from $J_\nu^{bc*}$, is 
%---------------
\begin{equation}
%---------------
\frac{\delta^{bc}}{\sqrt{2}}{\rm tr} \left( \Lambda_8^c T^a T^b T^a \right) 
= - \frac{C_F}{2}. 
%---------------
\end{equation}
%---------------
We note that the contributions in Eq.~(\ref{interference-frag}) contain
a phase $-1$ relative to the contributions in Eqs.~(\ref{unequal-frag})
and (\ref{exchange-frag}) that arises because of the interchange of the
identical quarks in an amplitude on only one side of the final-state
cut. We also note that the quantities $\sum_{i=1}^4 d_i(z, \lambda)$
[Eq.~(\ref{unequal-frag})], $\sum_{i=5}^8 d_i(z, \lambda)$
[Eq.~(\ref{exchange-frag})], and $\sum_{i=9}^{12} d_i(z, \lambda)$
[Eq.~(\ref{interference-frag})] are separately gauge invariant.

The dependence on $\bm{k}_{1\perp}$ in the expressions in
Eqs.~(\ref{exchange-frag}) and (\ref{interference-frag}) comes from the
quark- and gluon-propagator denominators, which contribute factors $[
\bm{k}_{1\perp}^2 + \left(\frac{2-z}{2z}M\right)^2]^{-1}$. Thus, the
integrals of these expressions over $\bm{k}_{1\perp}$ can be expressed
in terms of the scalar integrals $J_n [\left(\frac{2-z}{2z} M\right)^2]$
[Eq.~(\ref{Jn})]. Summing over the contributions in
Eqs.~(\ref{exchange-frag}) and (\ref{interference-frag}), using the
expressions for the absolute squares of the polarizations in
Eq.~(\ref{polarization-sums}), multiplying by the factor in
Eq.~(\ref{norm-factor}), and carrying out the phase-space
integration\footnote{We note that, although the final state contains two
identical quarks, the statistical factor $S$ in the phase space is
unity. This follows from the fact that there is no integration over the
momentum of the $Q\bar Q$ pair or from the fact that the two final-state
particles, namely, the $Q\bar Q$ pair and the single $Q$, are distinct.}
in Eq.~(\ref{2p-PS}), we obtain the following contributions to the
fragmentation functions:
%---------------
\begin{subequations}
\begin{eqnarray}
%---------------
\sum_{\lambda=0, \pm 1} \sum_{i=5}^{12}d_{i}(z,\lambda)
&=&
\frac{g_s^4 C_F (1-z)}{2 \pi N_c^2 z^{3+2 \epsilon} (2-z)^2 M} 
\Bigg\{ a_2^{[8]} J_2 \bigg[ \bigg( \frac{2-z}{2z} M \bigg)^2 \bigg]
\nonumber \\ && \hspace{4ex} 
+ M^2 a_3^{[8]} J_3 \bigg[ \bigg( \frac{2-z}{2z} M \bigg)^2 \bigg]
+ M^4 a_4^{[8]} J_4 \bigg[ \bigg( \frac{2-z}{2z} M \bigg)^2 \bigg] \Bigg\}, 
\\
{}
\sum_{i=5}^{12}d_{i}(z,\lambda=0)
&=&
\frac{g_s^4 C_F (1-z)^2}{2 \pi N_c^2 z^{5+2 \epsilon} (2-z)^2 M} 
\Bigg\{l_2^{[8]} J_2 \bigg[ \bigg( \frac{2-z}{2z} M \bigg)^2 \bigg]
\nonumber \\ && \hspace{4ex} 
+ M^2 l_3^{[8]} J_3 \bigg[ \bigg( \frac{2-z}{2z} M \bigg)^2 \bigg]
+ M^4 l_4^{[8]} J_4 \bigg[ \bigg( \frac{2-z}{2z} M \bigg)^2 \bigg]\Bigg\}, 
%---------------
\end{eqnarray}
\end{subequations}
%---------------
where the dimensionless coefficients $a_n^{[8]}$ and $l_n^{[8]}$ 
are given in terms of $z$ and $\epsilon$ by 
%---------------
\begin{subequations}
\begin{eqnarray}
%---------------
a_2^{[8]}&=& z^2 (z-1) [ -(9z^2+4z+4)
+ 2 \epsilon (3 z^2+4) 
+ \epsilon^2 (5 z^2-4 z-12) 
+ 2 \epsilon^3 (4-4 z+z^2)]
\nonumber \\
&& -2 N_c z (z-2) [ -2 (5 z^2-5 z+2) 
+ \epsilon (-z^3+ 8 z^2-6 z+4) 
+ \epsilon^2 ( z^3-2 z^2 ) ],
\\
{}a_3^{[8]} &=& - z (z-1) (z-2) 
[-2 (z^2+6 z-4) + \epsilon z (z+6) 
+ 2 \epsilon^2 z (z-2)]
\nonumber \\
&&
- 2 N_c z (z-1) (z-2)^2 (3-2 \epsilon) ,
\\
{}a_4^{[8]} &=& 
-(3-2 \epsilon) (z - 1)^2 (z - 2)^2, 
\\
{}l_2^{[8]} &=& z^4 [z+2+\epsilon (z-2)]^2
- 4 N_c z^3 (z-2) [ z+2+\epsilon(z-2)], 
\\
{}l_3^{[8]} &=& -z^2 (z-2) [ 
4 (z-1) (z+2) + \epsilon (z-2) (z^2+4 z-8)]
\nonumber \\
&& + 2 N_c z (z-2)^2 [ 4 (z-1) + \epsilon (z-2)^2], 
\\
{}l_4^{[8]} &=& 
2 [2 (z - 1) + \epsilon (z - 2)^2 ] (z - 1) (z - 2)^2.
%---------------
\end{eqnarray}
\end{subequations}
%---------------
Here, the terms that are proportional to $N_c$ come from the
contributions in Eq.~(\ref{interference-frag}). Such terms do not appear
in $a_4^{[8]}$ and $l_4^{[8]}$. Again, we retain the full $\epsilon$
dependence, as it may be useful for calculations of fragmentation
functions at higher orders in $\alpha_s$.

Taking the limit $\epsilon\to 0$, writing the result in the 
NRQCD-factorized form, and replacing free $Q\bar Q$ LDMEs with 
quarkonium LDMEs, we obtain
%---------------
\begin{subequations}
\label{equal-frag-eps-5-12}
\begin{eqnarray}
%---------------
\sum_{\lambda=0, \pm1} D_{Q\to H(\lambda)}^{(\textrm{5--12})} 
&=& 
\frac{\alpha_s^2 C_F (1-z)}{N_c^2 (2-z)^6 m_Q^3} 
\big[
z (1-z) (5 z^4-32 z^3+72 z^2-32 z+16) 
\nonumber\\
&&+8 N_c (2-z)^2 (z^3-6 z^2+6 z-2)
\big]\times 
\frac{\langle O^H({}^3S_1^{[8]}) \rangle}{3 (N_c^2-1)},
\label{equal-frag-eps-0-pol-sum}\\
D_{Q\to H(\lambda=0)}^{(\textrm{5--12})}
&=& 
\frac{\alpha_s^2 C_F
z (1-z)^2}
{3 N_c^2 (2-z)^6 m_Q^3} 
\big[ 
3 z^4-24 z^3+64 z^2-32 z+16 
\nonumber\\&&
+12 N_c (2-z)^2 (4-z)]
\frac{\langle O^H({}^3S_1^{[8]}) \rangle}{3 (N_c^2-1)}.
%---------------
\end{eqnarray}
\end{subequations}
%---------------
One obtains the complete fragmentation functions for the case in which
the initial quark $q$ and the quark $Q$ that is a constituent of
the quarkonium have the same flavor by adding the contributions in
Eq.~(\ref{equal-frag-eps-5-12}) to the contributions in 
Eq.~(\ref{ms-bar-frag-unequal}) with $r=m_q^2/(2m_Q)^2$ set equal to $1/4$.
The results are
%---------------
\begin{subequations}
\begin{eqnarray}
%---------------
&& 
\sum_\lambda D_{Q\to H(\lambda)}^{\overline{\rm MS}}
(z, \mu)
= 
\frac{\alpha_s^2 C_F}{2 z N_c^2 (2-z)^6 m_Q^3} 
\bigg[
N_c^2 ( z^2-2 z +2) (2-z)^6
\log \frac{\mu^2}{(2-z)^2 m_Q^2} 
\nonumber \\ 
&& \hspace{20ex} 
- N_c^2 z^2 (2-z)^4 (z^2-10z +10)
\nonumber \\ 
&& \hspace{20ex} 
+16 N_c z (2-z)^2 (1-z) (z^3-6 z^2+6 z-2)
\nonumber \\ 
&& \hspace{20ex} 
+ 2 z^2 (1-z)^2 (5 z^4-32 z^3+72 z^2-32 z+16) 
\bigg] 
\frac{\langle O^H({}^3S_1^{[8]}) \rangle}
{3 (N_c^2-1)}
, \label{equal-frag-eps-0}
\\
&& 
D_{Q\to H(\lambda=0)} (z) 
= 
\frac{\alpha_s^2 C_F(1-z)^2}{3 N_c^2 z (2-z)^6 m_Q^3} 
\bigg[
12 N_c^2 (2-z)^4 
+12 N_c z^2(2-z)^2 (4-z)
\nonumber \\ 
&& \hspace{20ex} + 
z^2(3 z^4-24 z^3+64 z^2-32 z+16 )
\bigg]
\frac{\langle O^H({}^3S_1^{[8]}) \rangle}{3 (N_c^2-1)}.
\label{equal-frag-eps-0-long-pol}
%---------------
\end{eqnarray}
\end{subequations}
%---------------
Identical expressions hold for the case of an initial antiquark. 

Our polarization-summed result in Eq.~(\ref{equal-frag-eps-0})
differs from the result in Eq.~(111) of
Ref.~\cite{Yuan:1994hn}, which was duplicated in Eq.~(A3) of
Ref.~\cite{Kniehl:1997gh}. In Ref.~\cite{Yuan:1994hn}, the color-octet
fragmentation function was obtained by multiplying the color-singlet
fragmentation function by a color factor. Consequently, the
contributions of the diagrams of Figs.~\ref{fig:unequal} and
\ref{fig:interference} were omitted. Our result in 
Eq.~(\ref{equal-frag-eps-0}) agrees with that in Eq.~(C29) of 
Ref.~\cite{Ma:2013yla}.

Our longitudinal-polarization result in
Eq.~(\ref{equal-frag-eps-0-long-pol}) agrees with the result in
Eq.~(C.16) of Ref.~\cite{hong-zhang}, once one takes into account the
fact that the LDME in Eq.~(\ref{equal-frag-eps-0-long-pol}) is 
a factor of three larger than the corresponding LDME in Ref.~\cite{hong-zhang}.

\section{Color-singlet fragmentation \label{sec:color-singlet}}

In this section, we compute the fragmentation function for a quark to
fragment through a spin-triplet color-singlet $S$-wave pair
$Q\bar{Q}(^3S_1^{[1]})$. This process proceeds at leading order in
$\alpha_s$ only if the fragmenting quark has the same flavor as the
quark in the $Q\bar Q$ pair. The Feynman diagrams that contribute to
this process at leading order in $\alpha_s$ are those in
Fig.~\ref{fig:exchange}. The corresponding contributions to the
fragmentation function for a quark to fragment into a
$Q\bar{Q}(^3S_1^{[1]})$ pair are identical to the contributions
$d_5$--$d_8$ in Eq.~(\ref{exchange-frag}), except that the color
projectors $\Lambda_8$ are replaced with color projectors 
$\Lambda_1$. In each contribution, the overall color factor is now
\begin{equation}
{\rm tr}(\Lambda_1T^aT^b\Lambda_1T^bT^a)=C_F^2, 
\label{singlet-CF}
\end{equation}
instead of $C_F/(2N_c)$ [Eq.~(\ref{exchange-CF})]. 

The dependence on $\bm{k}_{1\perp}$ in the expressions in
Eq.~(\ref{exchange-frag}) comes from the quark- and gluon-propagator
denominators, which contribute factors $[ \bm{k}_{1\perp}^2 +
\left(\frac{2-z}{2z} M\right)^2]^{-1}$. Thus, the integrals of these
expressions over $\bm{k}_{1\perp}$ can be expressed in terms of the
scalar integrals $J_n [\left(\frac{2-z}{2z} M\right)^2]$
[Eq.~(\ref{Jn})]. Summing over the contributions in
Eq.~(\ref{exchange-frag}), but taking the color factor in
Eq.~(\ref{singlet-CF}), using the expressions for the absolute squares
of the polarizations in Eq.~(\ref{polarization-sums}), multiplying by
the factor in Eq.~(\ref{norm-factor}), and carrying out the 
integration over the phase space in Eq.~(\ref{2p-PS}), 
we obtain the following contributions to the fragmentation
functions:
%---------------
\begin{subequations}
\begin{eqnarray}
%---------------
\sum_{\lambda=0, \pm 1} 
D_{Q\to Q\bar Q(^3S_1^{[1]})(\lambda)}^{(\textrm{5--8})}
&=&
\frac{ g_s^4 C_F^2(1-z)^2}{3 \pi z^{3+2 \epsilon} (2-z)^2 M} 
\biggl\{a_2^{[1]} J_2\left[\left(\tfrac{2-z}{2z} M\right)^2 \right] 
+ M^2 a_3^{[1]} J_3\left[\left(\tfrac{2-z}{2z} M\right)^2 \right]\nonumber\\
&&+ M^4 a_4^{[1]} J_4\left[\left(\tfrac{2-z}{2z} M\right)^2
\right] \biggr\}, 
\\ 
D_{Q\to Q\bar Q(^3S_1^{[1]})(\lambda=0)}^{(\textrm{5--8})} 
&=&
\frac{ g_s^4 C_F^2(1-z)^2}{3 \pi z^{5+2 \epsilon} (2-z)^2 M} 
\biggl\{l_2^{[1]} J_2\left[\left(\tfrac{2-z}{2z} M\right)^2 \right] 
+ M^2 l_3^{[1]} J_3\left[\left(\tfrac{2-z}{2z} M\right)^2 \right]\nonumber\\
&&+ M^4 l_4^{[1]} J_4\left[\left(\tfrac{2-z}{2z} M\right)^2 \right] \biggr\}, 
%---------------
\end{eqnarray}
\end{subequations}
%---------------
where  the dimensionless coefficients $a_n^{[1]}$ and $l_n^{[1]}$ 
are given in terms of $z$ and $\epsilon$ by 
%---------------
\begin{subequations}
\begin{eqnarray}
%---------------
a_2^{[1]} &=& z^2
[ (9 z^2 + 4 z + 4)
-\epsilon (6 z^2+8) 
- \epsilon^2 (5 z^2-4 z-12)
- 2 \epsilon^3 (z^2 - 4 z + 4) ], 
\\
{}a_3^{[1]} &=& 
z (z - 2) 
[- 2 (z^2 + 6 z - 4) + \epsilon z (2 \epsilon z - 4 \epsilon + z + 6) ] ,
\\
{}a_4^{[1]} &=& 
(3-2 \epsilon) (z - 1) (z - 2)^2, 
\\
{}l_2^{[1]} &=& [z + 2 + \epsilon (z - 2) ]^2 z^4,
\\
{}l_3^{[1]} &=& - (z-2) z^2 [4 (z + 2) (z - 1) + \epsilon (z^2 + 4 z - 8) (z - 2) ],
\\
{}l_4^{[1]} &=& 2 (z - 1) (z - 2)^2 [2 (z - 1) + \epsilon(z - 2)^2 ].
%---------------
\end{eqnarray}
\end{subequations}
%---------------

Taking the limit $\epsilon \to 0$, writing the result in the 
NRQCD-factorized form, and replacing free $Q\bar Q$ LDMEs with 
quarkonium LDMEs, we obtain
%---------------
\begin{subequations}
\begin{eqnarray}
%---------------
\sum_{\lambda=0, \pm1} D_{Q\to H(\lambda)}&=&
\frac{\alpha_s^2 C_F^2 z (1-z)^2 (5 z^4-32 z^3+72 z^2-32 z+16)
}{9 N_c (2-z)^6 m_Q^3}
\langle O^H({}^3S_1^{[1]}) \rangle,\label{sing-frag-sum}\\
D_{Q\to H(\lambda=0)}&=&
\frac{\alpha_s^2 C_F^2 z (1-z)^2 (3 z^4-24 z^3+64 z^2-32 z+16)}
{27 N_c (2-z)^6 m_Q^3}
\langle O^H({}^3S_1^{[1]}) \rangle,\label{sing-frag-long}
%---------------
\end{eqnarray}
\end{subequations}
%---------------
where the NRQCD LDME is defined by
\begin{equation}
\langle O^H({}^3S_1^{[1]}) \rangle=
\langle 0|
\chi^\dagger\sigma^i
\psi
\sum_{\lambda}{\cal P}_{H(P,\lambda)}
\psi^\dagger\sigma^i\chi|0\rangle,
\end{equation}
and, in writing the result in the NRQCD-factorized form, 
we have used the fact that
\begin{equation}
\langle O^{Q\bar{Q}({}^3S_1^{[1]})}({}^3S_1^{[1]}) \rangle
= 2N_c(d-1). 
\end{equation}

The result in Eq.~(\ref{sing-frag-sum}) agrees with the result in
Eq.~(16) of Ref.~\cite{Braaten:1993mp} and with the result in
Eq.~(C24) of Ref.~\cite{Ma:2013yla}, once one takes into account the
fact that the LDME in Eq.~(\ref{sing-frag-sum}) is a factor of $2N_c$
larger than the corresponding LDME in Ref.~\cite{Ma:2013yla}. Our
result in Eq.~(\ref{sing-frag-long}) agrees with the result in
Eq.~(C.11) of Ref.~\cite{hong-zhang}, once one takes into account the
fact that the LDME in  Eq.~(\ref{sing-frag-long}) is a factor of $6N_c$
larger than the LDME in  Ref.~\cite{hong-zhang}.

\section{Summary and discussion \label{sec:summary}}

In this paper we computed fragmentation functions for a quark $q$ to
fragment into a heavy quarkonium through a heavy $Q\bar Q$ channel in
which the $Q\bar Q$ pair is in a spin-triplet, $S$-wave state. Our
computations are at the leading nontrivial order in $\alpha_s$ and at
leading order in the heavy-quark velocity $v$. We have considered the
following cases: (i) the $q$ and $Q$ have different flavors, in which
case the $Q\bar Q$ pair must be in a color-octet state; (ii) the $q$ and
$Q$ have the same flavor and the $Q\bar Q$ pair is in a color-octet state;
(iii) the $q$ and $Q$ have the same flavor and the $Q\bar Q$ pair is in a
color-singlet state. In each case, we have computed both the
fragmentation function summed over all spin polarizations of the $Q\bar
Q$ pair and the fragmentation function for longitudinal polarization of
the $Q\bar Q$ pair. We have also presented expressions in
$d=4-2\epsilon$ dimensions, which may be useful in carrying out
calculations of fragmentation functions at higher orders in $\alpha_s$.

Our results for case~(i) are given for finite $m_q$ in
Eq.~(\ref{ms-bar-frag-unequal}) and for $m_q=0$ in
Eq.~(\ref{unequal-zero-mass}). The result in
Eq.~(\ref{ms-bar-frag-unequal-sum}) for the sum over $Q\bar Q$
polarizations agrees with previous calculations in
Refs.~\cite{Ma:1995vi,Ma:2013yla}, and the result in
Eq.~(\ref{unequal-zero-mass-sum-pol}) for the sum over $Q\bar Q$
polarizations agrees with a calculation of quark fragmentation into
lepton pairs in Ref.~\cite{Braaten:2001sz}, once differences in the
phase-space and color factors have been taken into account. The
result in Eq.~(\ref{ms-bar-frag-unequal-long}) corrects the
result in Ref.~\cite{hong-zhang}. This correction has been confirmed by 
the author of Ref.~\cite{hong-zhang}.

Our result for case~(ii) for the fragmentation function summed over
$Q\bar Q$ polarizations is given in 
Eq.~(\ref{equal-frag-eps-0}) and agrees with the result in
Ref.~\cite{Ma:2013yla}, but disagrees with the result in 
Ref.~\cite{Yuan:1994hn}. %XXXXXXX
Our result for case~(ii) for the
longitudinal-polarization fragmentation function is given in
Eq.~(\ref{equal-frag-eps-0-long-pol}) and agrees with the
result in Ref.~\cite{hong-zhang}.

Our result for case~(iii) for the fragmentation function summed over
$Q\bar Q$ polarizations is given in Eq.~(\ref{sing-frag-sum}) and agrees
with the results in Refs.~\cite{Braaten:1993mp,Ma:2013yla}. Our result
for case~(iii) for the longitudinal-polarization fragmentation
function is given in Eq.~(\ref{sing-frag-long}) and agrees with the
result in Ref.~\cite{hong-zhang}.

The new results for longitudinally polarized fragmentation functions
that we have obtained in this paper will make it possible to compute
quark-initiated leading-power fragmentation contributions to the
production of polarized $S$-wave spin-triplet quarkonia. While 
gluon-initiated fragmentation dominates quark-initiated 
fragmentation in quarkonium hadroproduction, quark-initiated 
fragmentation may be important for other quarkonium production 
processes, such as production in $e^+e^-$ annihilation.

As we have mentioned, our calculations are at the leading nontrivial
order in $\alpha_s$ and $v$. A complete calculation at order
$\alpha_s^5$ of the LP contributions to quarkonium production in the
color-octet channels requires corrections to the fragmentation functions
for the $^1S_0$ and $^3P_J$ channels at next-to-leading order (NLO) in
$\alpha_s$ and for the $^3S_1$ channel through next-to-next-to-leading
order in $\alpha_s$. Corrections of NLO in $\alpha_s$ to the
fragmentation function for a gluon to fragment into a quarkonium through
the $^3S_1$ color-octet channel have already been computed
\cite{Braaten:2000pc,Ma:2013yla} and give a contribution that is
numerically large in comparison with the leading-order (LO) contribution
\cite{Bodwin:2014gia}. Corrections of higher order in $v$ are also known
to be large, relative to the LO contribution, for the fragmentation
functions for gluons to fragment into quarkonia through the $^3S_1$
color-singlet and color-octet channels
\cite{Bodwin:2003wh,Bodwin:2012xc}. These large corrections of higher
order in $\alpha_s$ and $v$ suggest that it may be important to compute
higher-order corrections for additional quarkonium production channels
and for quark-initiated, as well as gluon-initiated, fragmentation
processes.

%--------------------------------------------------------------------
\begin{acknowledgments}
We thank Jianwei Qiu for pointing out Ref.~\cite{hong-zhang} to us. We are
grateful to Hong Zhang for helpful discussions regarding comparisons
between our work and the results in Refs.~\cite{Ma:2013yla, hong-zhang}.
The work of G.T.B.\ and H.S.C.\ is supported by the U.S.\ Department of
Energy, Division of High Energy Physics, under Contract No.\ 
DE-AC02-06CH11357. The work of U-R.K.\ is supported by the National
Research Foundation of Korea under Contract No.\ NRF-2012R1A1A2008983. 
The submitted manuscript has been created in part by UChicago Argonne, LLC,
Operator of Argonne National Laboratory. Argonne, a U.S.\ Department of
Energy Office of Science laboratory, is operated under Contract No.\
DE-AC02-06CH11357. 

\end{acknowledgments}
\vskip 5ex
{\it Note added.---} After the first version of this paper was submitted
to the arXiv, Ref.~\cite{Ma:2015yka} appeared. That paper confirms our
results.

\end{document}